\documentclass[12pt,fleqn,twoside]{article}
\usepackage{graphicx}
\usepackage[small,bf]{caption2}
\usepackage{latexsym}
\usepackage{amsfonts}
\usepackage{amssymb}
\usepackage{amsmath}

\newcommand{\bea}{\begin{eqnarray}}
\newcommand{\eea}{\end{eqnarray}}
\newcommand{\bef}{\begin{figure}}
\newcommand{\enf}{\end{figure}}
\newcommand{\ball}{\begin{array}{ll}}
\newcommand{\bal}{\begin{array}{l}}
\newcommand{\ea}{\end{array}}
\newcommand{\feta}{{\mbox{\boldmath$\eta$\unboldmath}}}

\newcommand{\rel}{{\mathbb{R}}}
\newcommand{\nat}{{\mathbb{N}}}
\newcommand{\ganz}{{\mathbb{Z}}}
\newcommand{\ord}{{\mathcal{O}}}

\newcommand{\suxl}{\sum_{x=1}^{L}}
\newcommand{\proxl}{\prod_{x=1}^{L}}

\newcommand{\xbo}{{\mathbf{x}}}

\numberwithin{equation}{section}

\title{Condensation in the zero range process: stationary and dynamical properties}
\author{Stefan Gro\ss kinsky$^1$, Gunter M.\ Sch\"utz$^2$, and Herbert Spohn$^1$}
\date{\today}

\begin{document}
\maketitle

\renewcommand{\thefootnote}{\arabic{footnote}}
\footnotetext[1]{Zentrum Mathematik, Technische Universit\"at M\"unchen, 85747 Garching bei M\"unchen, Germany; e-mail: stefang@ma.tum.de, spohn@ma.tum.de}
\footnotetext[2]{Institut f\"ur Festk\"orperforschung, Forschungszentrum J\"ulich, 52425 J\"ulich, Germany; e-mail: g.schuetz@fz-juelich.de}

\begin{abstract}
The zero range process is of particular importance as a generic model for domain wall dynamics of one-dimensional systems far from equilibrium. We study this process in one dimension with rates which induce an effective attraction between particles. We rigorously prove that for the stationary probability measure there is a background phase at some critical density and for large system size essentially all excess particles accumulate at a \textit{single}, randomly located site. Using random walk arguments supported by Monte Carlo simulations, we also study the dynamics of the clustering process with particular attention to the difference between symmetric and asymmetric jump rates. For the late stage of the clustering we derive an effective master equation, governing the occupation number at clustering sites.
\end{abstract}

\section{Introduction}
Low dimensional stochastic particle systems far from equilibrium have a much richer structure than their equilibrium counterparts. In particular, even in one spatial dimension there is the possibility of a phase transition. On the other hand, we do not have available general criteria which would allow us to determine, for example, the phase diagram. A recent attempt in this direction is a proposal by Kafri et al.\ \cite{kafrietal02}, who study phase separation in one dimension. Roughly speaking, they map the domain wall dynamics of clusters to a zero range process, for which analytical tools are available. Under suitable conditions, the clusters tend to grow. This coarsening is an intriguing phenomenon already on the level of the zero range process itself, and we will investigate it in much greater detail than available so far.

The zero range process is a stochastic particle system on the lattice $\mathbb{Z}^d$ where the jump rate $g(k)$ of a given particle depends only on the occupation number $k$ at its current position. This model was originally introduced as a simple example of an interacting Markov process \cite{spitzer70}. Various properties have been established, among them the existence of the dynamics under very general conditions, classification of invariant measures, and hydrodynamic limits \cite{andjel82,kipnislandim,liggettetal81}. If $g(k)$ is decreasing in $k$, then this induces an effective attraction between particles, as first noted in \cite{bialasetal97,evans00}. As a result, there is a critical background density and excess particles condense on a non-extensive fraction of the volume.

Such a clustering phenomenon can be studied on two distinct levels. Firstly, the phenomenon is present already in the steady state. Based on some results for large deviations of independent, identically distributed random variables \cite{vinogradov,baltrunasetal02} and under general assumptions on $g(k)$, we will prove that for a typical steady state configuration there is a background phase at some critical density. Any additional mass is concentrated at a \textit{single}, randomly located site. For the model introduced in \cite{evans00}, we will analyze the statistical properties of the background phase in detail. Secondly, there is the dynamics of clustering with the steady state appearing only in the long time limit. In a very recent preprint Godr\`eche \cite{godreche03} addresses this problem. He assumes a uniform initial density and investigates numerically how the probability distribution of the number of particles at some given site evolves in time, with particular attention to the macroscopic component of that distribution. In view of our static result we study the dynamics of Evans' model \cite{evans00}, pursuing a somewhat different approach: During the initial nucleation process random sites are selected, at which a macroscopic number of particles accumulates. We investigate the effective dynamics of the number of particles at such cluster sites, in particular how the smaller occupation numbers become extinguished to the benefit of the larger ones.

Our contribution should be understood as a case study in the context of phase transitions in one-dimensional systems far from equilibrium, which has been a topic of major interest in the past decade (see \cite{schuetz00,mukamel00} and references therein). Of particular interest is the occurrence of phase separation in systems with two conservation laws (see e.g.\ \cite{kafrietal03} and \cite{evansetal98,arndtetal98}) whose macroscopic behavior has only recently been examined to some extent \cite{tothetal02,grosskinskyetal03,popkovetal02}. Given the correspondence to the zero-range process our results also provide new information on the stability of domain walls (shocks) which separate macroscopic regions of different phases in two-component systems. Domain wall stability already proved to be a key ingredient in the theory of boundary-induced phase transitions in systems with one conservation law \cite{krug91,kolomeiskyetal98,popkovetal99} and thus may shed light on boundary-induced spontaneous symmetry breaking \cite{evansetal95} in two-component systems.

To give a brief outline: in the following Section we discuss the stationary measures for the zero range process and in Section 3 the two main results on the equivalence of ensembles and the structure of the condensed phase are established. In Sections 4 and 5 we study the condensation transition for the model introduced in \cite{evans00}, first on the level of the stationary measure, and secondly through the dynamics of the clustering.

\section{Zero range process and its infinite volume stationary measures}
For notational simplicity we restrict ourselves to a description in one space dimension, but our results on the equivalence of ensembles hold for arbitrary dimension. Rather than defining the zero range process directly on an infinite lattice (cf.\ \cite{andjel82,liggettetal81}), we first consider a finite system, compute the (unique) stationary measure and analyze it in the limit of infinite system size.

Let us consider a zero range process on the one-dimensional lattice $\Lambda_L =\{ 1,\ldots ,L\}$ of $L$ sites with periodic boundary conditions. Let $\eta_x \in\nat$ be the number of particles on site $x\in\Lambda_L$, where $\nat =\{ 0,1,2,\ldots \}$. The state space is given by $\Omega_L =\nat^{\Lambda_L}$ and we denote a particle configuration by $\feta =(\eta_x )_{x\in\Lambda_L} \in\Omega_L$. At a given site $x\in\Lambda_L$, the number $\eta_x$ of particles decreases by one after an exponential waiting time with rate $g(\eta_x )$ and the leaving particle jumps to site $x+y$ with probability $p(y)$. The jump probabilities $p:\ganz\to [0,1]$ are normalized, $\sum_y p(y)=1$, $p(0)=0$, assumed to be of finite range, $p(y)=0$ for $|y|>R$, and irreducible, $p(1)>0$. For the dynamics to be well defined in the limit $L\to\infty$ and to be nondegenerate the rate function $g:\nat \to [0,\infty )$ has to satisfy
\bea\label{gassum}
\sup_{k\in\nat} |g(k+1)-g(k)|<\infty\ ,\quad g(k)>g(0)=0\ \mbox{for all }k>0.
\eea
The generator of the zero range process is then given by
\bea\label{generator}
(\mathcal{L} f)(\feta )=\suxl\sum_{y=-R}^R g(\eta_x )\, p(y)\big( f(\feta^{x,x+y} ) -f(\feta )\big) ,
\eea
regarded as a linear operator on $C(\Omega_L ,\rel )$. We used the shorthand $\eta_z^{x,x+y}=\eta_z -\delta (z,x) +\delta\big( z,((x+y-1)mod\, L)+1\big)$ for all $z\in\Lambda_L$, where $\delta (.,.)$ denotes the Kronecker delta function. The process conserves the number of particles $\Sigma_L (\feta ) =\suxl\eta_x$, thereby dividing the configuration space into the finite, invariant subsets $\Omega_{L,N} =\{\feta\in\Omega_L |\Sigma_L (\feta )=N\}$ with $N\in\nat$. Restricted to $C(\Omega_{L,N} ,\rel )$ with $L,N$ fixed, $\mathcal{L}$ is a finite dimensional matrix and the process is well defined. However for $L=\infty$ this is true only for ``reasonable'' initial conditions and under the assumption (\ref{gassum}), see \cite{andjel82,liggettetal81}.

The following results for the stationary measures are well known and taken from \cite{spitzer70,andjel82,evans00}. The zero range process (\ref{generator}) on $\Omega_{N,L}$ is an irreducible Markov jump process with the unique stationary measure
\bea\label{canens}
\mu^{N,L} (\feta )={1\over Z(N,L)}\proxl W(\eta_x )\,\delta\!\left(\Sigma_L (\feta ),N\right) .
\eea
The weight $W$ is given by
\bea
W(k):=\prod_{i=1}^k {1\over g(i)}
\eea
and the normalizing partition function is
\bea\label{canz}
Z(N,L)=\sum_{\feta\in\Omega_L}\proxl W(\eta_x )\,\delta\!\left(\Sigma_L (\feta ),N\right) .
\eea
Clearly (\ref{canens}) resembles a canonical ensemble in statistical mechanics. Therefore, in the limit of large system size $L,N\to\infty$ with fixed particle density $\rho =N/L$, (\ref{canens}) is expected to be equivalent to a grand canonical product measure, which is defined through
\bea\label{grandcanens}
\nu_\phi^L (\feta )=\proxl\nu_\phi (\eta_x )\quad\mbox{with }\nu_\phi (k)={1\over Z(\phi )}\, W(k)\phi^k \ ,
\eea
and where the fugacity $\phi\geq 0$ is adjusted to fix the average density.

Let $\phi_c$ be the radius of convergence of the grand canonical (one site) partition function
\bea\label{grandcanz}
Z(\phi )=\sum_{k=0}^\infty W(k)\,\phi^k .
\eea
The measure (\ref{grandcanens}) is well defined for fugacities $\phi\in [0,\phi_c )$ and its average particle density $\rho (\phi )$ as a function of $\phi$ is given by
\bea\label{rhodef}
\rho (\phi )=\sum_{k=0}^\infty k\,\nu_\phi (k)=\phi {\partial \log Z(\phi )\over\partial\phi }\ .
\eea
The range of $\rho$ is the interval $[0,\rho_c )$, with $\rho (0)=0$ and $\rho_c =\lim_{\phi\nearrow\phi_c} \rho (\phi )$ the critical density. $\phi\to\rho (\phi )$ is strictly increasing and we denote the inverse function on $[0,\rho_c )$ by $\phi (\rho )$. If $\phi_c =\infty$, then $\rho_c =\infty$ (see \cite{kipnislandim}, Lemma 2.3.3), whereas for $\phi_c <\infty$, both $\rho_c =\infty$ and $\rho_c <\infty$ are possible. In the second case $Z(\phi_c )<\infty$ (see \cite{kipnislandim}, Lemma 2.3.3) and $\nu_{\phi_c}$ is a well defined probability measure with $\langle\eta_x \rangle_{\nu_{\phi_c}}=\rho_c$. Thus we set
\bea\label{phirho}
\phi (\rho )=\left\{\ball\mbox{inverse of }\rho (\phi )\ &,\quad\mbox{for }\rho <\rho_c\\ \qquad\phi_c &,\quad\mbox{for }\rho\geq\rho_c \ea\right.\ ,
\eea
where $\rho_c$ can be either finite or infinite, cf.\ Figure 1 in Section 4. The reason for this particular convention will become clear in the next section. In this way we may regard the measure $\nu_\phi$ also as a function of $\rho$ through $\nu_{\phi (\rho )}$.

The link between canonical and grand canonical measures is given through the pointwise limit of the $n$-point marginal,
\bea
\lim\limits_{L\to\infty}\mu_n^{[\rho L] ,L} (\mathbf{k})=\prod\limits_{i=1}^n \nu_{\phi (\rho )} (k_i ).
\eea
Here $[a]$ denotes the integer part of $a\in\rel$ and for every $n\in\nat$, $\mathbf{x}=(x_1 ,\ldots ,x_n )\in\Lambda_L^n$ with $x_i \neq x_j$ for $i\neq j$, $\mathbf{k}=(k_1 ,\ldots ,k_n )\in\nat^n$ the $n$-point marginal is defined as $\mu_{n,\xbo}^{N,L} (\mathbf{k}):=\mu^{N,L} (\{\eta_{x_1}=k_1 ,\ldots ,\eta_{x_n}=k_n \} )$. Since the measure $\mu^{N,L}$ is permutation invariant (see (\ref{canens})), the marginals do not depend on the sites $\xbo$ individually, but only on their number $n$. For this reason only $n$ is specified in our notation.

A rigorous result on this equivalence is available in \cite{kipnislandim}, Appendix 2, but it does not cover the supercritical case $\rho_c <\infty$ and $N/L\geq\rho_c$. It will be discussed in Section 3, where particular attention is given to the statistical properties of the excess density. If $\rho_c <\infty$, then $\nu_{\phi_c}$ is well defined and by necessity decays subexponentially, which will be explained in (\ref{subexp}). Examples with power-law decay are given in \cite{bialasetal97}, \cite{kipnislandim} (Example 2.3.4). In \cite{evans00} Evans studies the dependence on the power-law exponent in more detail and introduces a generic model where $\rho_c$ can be either finite or infinite, depending on a system parameter. Its stationary and dynamical properties will be studied in detail in Sections 4 and 5.

The jump probabilities $p$ do not influence the stationary measures, but they play an important role for the relaxation dynamics, see Section 5. The zero range process is reversible if and only if $p$ is symmetric. The stationary current is given by $j=m(p)\langle g(\eta_x )\rangle$, where $m(p)=\sum_{y=-R}^R yp(y)$ is the first moment of $p$, which for non-symmetric jump probabilities is generically non-zero. The average jump rate $\langle g\rangle$ of the two ensembles is given by
\bea\label{current}
\langle g(\eta_x )\rangle_{\mu^{N,L}}= {Z(N-1,L)\over Z(N,L)}&\quad&\mbox{(canonical)},\nonumber\\
\langle g(\eta_x )\rangle_{\nu_\phi^L}=\sum_{k=0}^\infty g(k)\nu_\phi (k)=\phi&\quad&\mbox{(grand-canonical)}.
\eea
Thus for $m(p)>0$ the grand canonical current $j(\rho )=m(p)\phi (\rho )$ is monotone increasing in $\rho$, approaching its maximum value $m(p) \phi_c$ as $\rho\to\rho_c$, and correspondingly for $m(p)<0$.

\section{Equivalence of supercritical measures}
We consider the zero range process with $\rho_c <\infty$ and supercritical canonical measures with $\rho =N/L>\rho_c$. The heuristic picture, developed in \cite{bialasetal97} and \cite{evans00}, is that most sites of the system are distributed according to $\nu_{\phi_c}$ with mean occupation number $\rho_c$. For large $L$, the $(\rho -\rho_c )L$ excess particles presumably condense on a few sites. If so, locally one will observe the grand-canonical ensemble with $\phi =\phi_c$. This picture is made precise in\\

\textbf{Theorem 1.} (Equivalence of ensembles)\\
Let $\rho\to\phi (\rho )$ be defined as in (\ref{rhodef}), (\ref{phirho}). Then for every $\rho\in [0,\infty )$ the $n$-th marginal has the pointwise limit
\bea\label{equivens2}
\lim_{L\to\infty} \mu_n^{[\rho L] ,L} (\mathbf{k})=\prod_{i=1}^n \nu_{\phi (\rho )} (k_i )\ .
\eea
The canonical partition functions converge as
\bea\label{equivz2}
\lim_{L\to\infty} {1\over L}\log Z([\rho L] ,L)=\log Z(\phi (\rho )) -\rho\log\phi (\rho )\ .
\eea\\

\textit{Proof.} It is convenient to characterize the distance between the canonical and grand-canonical measures through the relative entropy $S$. For two arbitrary probability measures $\mu ,\nu$ on a countable set $\Omega$ it is defined as
\bea\label{entrdef}
S(\mu |\nu )=\left\{\ball\sum\limits_{\omega\in\Omega}\mu (\omega )\log {\mu (\omega )\over\nu (\omega)}\ &,\quad\mbox{if }\mu\ll\nu\\
\qquad\quad\ \infty\ &,\quad\mbox{otherwise}\ea\right.
\eea
where $\mu\ll\nu$ means that $\mu$ is absolutely continuous w.r.t.\ $\nu$. It has the properties (see e.g.\ \cite{wehrl78})
\bea\label{pos}
S(\mu |\nu )\geq 0\ ,\quad S(\mu |\nu )=0 \Leftrightarrow \mu (\omega )=\nu (\omega )\mbox{ for all }\omega\in\Omega\ .
\eea
To prove (\ref{equivens2}) it is therefore enough to establish that
\bea\label{toshow}
\lim_{L\to\infty} S\left(\mu_n^{[\rho L] ,L}\,\big|\,\nu_{\phi (\rho )}^n\right) =0
\eea
for every $\rho\in [0,\infty )$.

We recall that $\Sigma_L (\feta )=\suxl\eta_x$ denotes the number of particles. By definition (\ref{canens}) one has for a fixed $\feta\in\Omega_L$
\bea\label{zerg}
\mu^{N,L} (\feta )={W^L (\feta )\,\delta\big(\Sigma_L (\feta ),N\big)\over Z(N,L) }\ ,
\eea
where $W^L$ denotes the product measure $W^L (\feta )=\prod_{x=1}^L W(\eta_x )$. Thus, using (\ref{entrdef}), it follows that for every $\phi\in [0,\phi_c ]$
\bea\label{erg1}
S\left(\mu^{N,L}\,\big|\,\nu_\phi^L \right) =\sum_{\feta \in\Omega_{L,N}}\mu^{N,L}(\feta )\,\log {W^L (\feta )\over\nu_\phi^L (\feta )\, Z(N,L) }\ ,
\eea
since $\mu^{N,L}$ is absolutely continuous w.r.t.\ $\nu_\phi^L$. From (\ref{canz}) and (\ref{grandcanens}) we conclude that
\bea\label{zrel}
\nu_\phi^L (\feta )\, Z(\phi )^L &=&W^L (\feta )\, \phi^N \qquad\mbox{for }\feta\in\Omega_{L,N}\quad\mbox{and}\nonumber\\
\nu_\phi^L \big(\{\Sigma_L =N\}\big)\, Z(\phi )^L &=&Z(N,L)\,\phi^N \ .
\eea
To simplify notation we use the shorthand $\nu_\phi^L (A)=\sum_{\feta\in A} \nu_\phi^L (\feta )$ for a subset $A$ of the configuration space and $\{\feta\, |\,\Sigma_L (\feta )=N\} =\{\Sigma_L =N\}$. Inserting in (\ref{erg1}) this yields
\bea
S\left(\mu^{N,L}\,\big|\,\nu_\phi^L \right) =-\log\nu_\phi^L \left(\{\Sigma_L =N\} \right) .
\eea
At this point we use the subadditivity of $S$, namely if two measures $\mu$, $\nu$ have marginals $\mu_i$, $\nu_i$, $i=1,2$, then
\bea\label{sub}
S(\mu |\nu )\geq S(\mu_1 |\nu_1 )+S(\mu_2 |\nu_2 )\ .
\eea
Therefore for every $n\in\{ 1,\ldots ,L\}$ and $\phi\in [0,\phi_c ]$
\bea\label{erg2}
S\left(\mu_n^{N,L}\,\big|\,\nu_\phi^n \right)\leq-{1\over
  [L/n]}\log\nu_\phi^L \left(\{\Sigma_L =N\} \right).
\eea

The key point is to maximize $\nu_\phi^L \left(\{\Sigma_L =N\}\right)$ by appropriately adjusting $\phi =\phi (\rho )$ as defined in (\ref{phirho}). In the subcritical case, $\rho =N/L<\rho_c$, we have $\phi (\rho )<\phi_c$ and $\nu_{\phi (\rho )}$ has exponential moments (see (\ref{grandcanens})). Then the variance $\sigma^2$ of $\nu_\phi$ is finite and the limit distribution of $(\Sigma_L -(\rho L))/(\sigma\sqrt{L})$ is given by the normal distribution $\mathcal{N} (0,1)$ (cf.\ (\ref{clt})). By the local limit theorem (see e.g.\ \cite{gnedenko}) we get in this case $\nu_{\phi (\rho )}^L \left(\{\Sigma_L =[\rho L]\} \right)\simeq 1/\sqrt{L}$ for large $L$. For $\rho =\rho_c$ the decay of $\nu_{\phi_c}$ is subexponential, since $\phi =\phi_c$ is the radius of convergence of the partition function $Z(\phi )$ in (\ref{grandcanz}). Thus
\bea\label{subexp}
\lim_{k\to\infty}{1\over k} \log \nu_{\phi_c}(k )=0
\eea
and the second moment of $\nu_{\phi_c}$ could be infinite, leading to a non-normal limit distribution (cf.\ (\ref{cltlevy})). Since the first moment of $\nu_{\phi_c}$ equals $\rho_c <\infty$, by the local limit theorem for non-normal limit distributions (see also \cite{gnedenko}) we get the lower bound $\nu_{\phi_c}^L \left(\{\Sigma_L =[\rho_c L]\} \right)\gtrsim 1/L$. The supercritical case $\rho >\rho_c$, where $\phi (\rho )=\phi_c$, can be reduced to the critical one via
\bea\label{reduce}
\lefteqn{\nu_{\phi_c}^L \Big(\big\{\Sigma_L =[\rho L]\big\}\Big)\geq }\nonumber\\
& &\qquad\geq\nu_{\phi_c}^L \Big(\big\{\eta_L =[\rho L]-[\rho_c (L-1)],\Sigma_{L-1} =[\rho_c (L-1)] \big\}\Big)=\nonumber\\
& &\qquad=\nu_{\phi_c} \Big( [\rho L]-[\rho_c (L-1)]\Big)\nu_{\phi_c}^{L-1} \Big(\big\{\Sigma_{L-1} =[\rho_c (L-1)]\big\}\Big)\ .
\eea
Both terms decay subexponentially, the first one using (\ref{subexp}) and the second one as in the critical case. Thus in all cases we have a subexponential lower bound on $\nu_{\phi (\rho )}^L \left(\{\Sigma_L =[\rho L]\}\right)$ and the limit (\ref{toshow}) follows for all $\rho\in [0,\infty )$ from (\ref{erg2}),
\bea
\lim_{L\to\infty}\! S\Big(\mu_n^{[\rho L] ,L}\,\big|\,\nu_{\phi (\rho
  )}^n\!\Big)\leq -\!\!\lim_{L\to\infty}\! {1\over [L/n]}\log
\nu_{\phi (\rho )}^L \!\big(\{\Sigma_L \! =\! [\rho L]\}\big) =0\, .
\eea

To establish (\ref{equivz2}) we use the second line of (\ref{zrel}) and immediately get
\bea
{1\over L}\log Z(N,L)-{1\over L}\log \nu_\phi^L \left(\{\Sigma_L =N\}\right) =\log Z(\phi ) -{N\over L}\log\phi\ ,
\eea
for all $N,L$. With $\phi =\phi (N/L)$ we can use the above estimates, so that the second term on the left vanishes in the limit $L\to\infty$ and the assertion (\ref{equivz2}) follows.\hfill $\Box$\\
\\
Theorem 1 ensures us that the volume fraction of the condensed phase vanishes in the limit $L\to\infty$. In principle it could still contain an infinite number of sites, and the question remains, how many condensed sites there are in a typical configuration. The answer depends on the large-$k$ behavior of the critical distribution $\nu_{\phi_c} (k)$. From (\ref{subexp}) we know already that it decays subexponentially. We will show that for a large class of such subexponential distributions the excess particles condense on a single, randomly located site.

Let $\mathcal{C}_2$ be the set of distributions $\nu_{\phi_c}$ which have finite second moment and for which the integrated tail $\nu_{\phi_c} \big(\{\eta_1 \geq k\}\big)$ is heavier than $\exp [-k^\alpha ]$ for some $\alpha\in (0,1/2)$. $\mathcal{C}_p$ denotes the set of distributions with power-law tail $\nu_{\phi_c} (k)\simeq k^{-b}$, $b\in (2,3]$, for which the second moment diverges. For a detailed description and results on subexponential distributions we refer the reader to \cite{baltrunasetal02,goldieetal98}.\\

\textbf{Theorem 2.} Let $\nu_{\phi_c}$ be in the class $\mathcal{C}_2 \cup\mathcal{C}_p$ as defined above, with first moment $\rho_c <\infty$. Then for the sequence of the corresponding canonical measures $\mu^{[\rho L],L}$ with $\rho >\rho_c$ one has
\bea\label{cor}
\lim_{L\to\infty} \mu^{[\rho L],L} \left(\Big\{\max\limits_{1\leq x\leq L} \eta_x \geq [(\rho -\rho_c )L]\Big\}\right) =1.
\eea\\[-2mm]

The probability that there is a site which contains at least $(\rho -\rho_c )L$ particles converges to one in the thermodynamic limit. But the occupation number cannot be substantially larger than $(\rho -\rho_c )L$, since almost all sites are distributed according to $\nu_{\phi_c}$, as proved in Theorem 1. Therefore all excess particles are condensed on a single site in the limit $L\to\infty$. The proof of Theorem 2 uses large deviation results on the asymptotic behavior of $\nu_{\phi_c}^L \big(\{\Sigma_L \geq\rho L\}\big)$ for $L\to\infty$, which we summarize in the following Lemma for our purpose.\\

\textbf{Lemma.} Let $\eta_1 ,\eta_2 ,\ldots$ be i.i.d.\ random variables with mean zero and probability distribution $P\in\mathcal{C}_2 \cup\mathcal{C}_p$. Then with $\Sigma_L =\suxl \eta_x$ one has for any $\rho >0$
\bea\label{ldresult}
\lim_{L\to\infty} \left| {P^L \big(\{\Sigma_L \geq\rho L\}\big)\over L\, P\big(\{\eta_1 \geq\rho L\}\big)} -1\right| =0\ .
\eea\\[-2mm]

\textit{Proof of the Lemma.} In the case $P\in \mathcal{C}_p$ see \cite{vinogradov} (Chapter 1, Corollary 1.1.1 to 1.1.3) and for distributions in $\mathcal{C}_2$ see \cite{baltrunasetal02}.\\

The interpretation of the Lemma is that under a distribution in $\mathcal{C}_2 \cup\mathcal{C}_p$ the rare event $\{ \Sigma_L \geq \rho L\}$ in the limit $L\to\infty$ is realized by the deviation of a single (randomly positioned) site with probability one. Thus for $\nu_{\phi_c} \in\mathcal{C}_2 \cup\mathcal{C}_p$ this is also a typical configuration under the canonical measure $\mu^{[\rho L],L}$, since for $\rho >\rho_c$ the latter is basically given by the grand canonical critical measure $\nu_{\phi_c}$ under the condition $\{\Sigma_L = [\rho L]\}$, cf.\ (\ref{umr}). This argument will be made precise in the following.\\

\textit{Proof of Theorem 2.} Obviously one has
\bea
\mu^{[\rho L],L} \left(\Big\{\max\limits_{1\leq x\leq L} \eta_x \geq [(\rho -\rho_c )L]\Big\}\right) \leq 1
\eea
for all $L\in\nat$. To find a lower bound that converges to $1$ for $L\to\infty$ we apply the results of the Lemma. We note that for all $\feta\in\Omega_{L,N}$ it is
\bea\label{umr}
\mu^{[\rho L],L} \!\Big(\!\Big\{\!\max\limits_{1\leq x\leq L} \eta_x \!\geq\! [(\rho -\rho_c )L]\Big\}\!\Big)\! =\! {\nu_{\phi_c}^L \!\Big(\!\Big\{\!\max\limits_{1\leq x\leq L} \eta_x \!\geq\! [(\rho -\rho_c )L]\Big\}\!\Big)\over \nu_{\phi_c}^L \big(\{\Sigma_L =[\rho L]\}\big)}\
\eea
cf.\ (\ref{zerg}), and for $L\to\infty$ 
\bea\label{lmax}
\nu_{\phi_c}^L \Big(\!\Big\{\!\max\limits_{1\leq x\leq L} \eta_x \!\geq\! [(\rho\! -\rho_c \! )L]\Big\}\!\Big)\! =\! L\,\nu_{\phi_c}\! \big(\!\big\{\eta_1 \!\geq\! [(\rho\! -\rho_c \! )L]\big\}\!\big) (1\! +\! o(1))\, .
\eea
Shifting the expectation value to zero via $\eta_x '=\eta_x -\rho_c$, $\Sigma '_L :=\suxl \eta_x '$ and using $\nu_{\phi_c}^L \big(\{\Sigma_L =[\rho L]\}\big)\leq \nu_{\phi_c}^L \big(\{\Sigma_L \geq [\rho L]\}\big)$ we get from (\ref{umr}) and (\ref{lmax})
\bea\label{fertig}
\lefteqn{\mu^{[\rho L],L} \Big(\Big\{\max\limits_{1\leq x\leq L} \eta_x \geq [(\rho -\rho_c )L]\Big\}\Big)\geq }\nonumber\\
& &\qquad\qquad\qquad\geq {L\,\nu_{\phi_c} \big(\big\{\eta_1 ' \geq [(\rho -\rho_c )L]\big\}\big)\over \nu_{\phi_c}^L \big(\big\{\Sigma_L ' \geq [(\rho -\rho_c )L]\big\}\big) }\big( 1\! +\! o(1)\big)\ .
\eea
By (\ref{ldresult}) the righthand side of (\ref{fertig}) converges to $1$ in the limit $L\to\infty$ and (\ref{cor}) is shown.\hfill$\Box$\\

\section{Stationary properties near criticality}
In \cite{evans00} Evans studies the zero range process with rates 
\bea
g_b (k)=\theta (k)(1+b/k)\ ,
\eea
where $\theta (0)=0$ and $\theta (k)=1$ for $k>0$. He observes that for $b>2$ the critical density $\rho_c <\infty$. Our goal here is to study the properties of the invariant measures for any $b>0$. The stationary weight for the zero range process with rates $g_b$ is given by
\bea\label{weights}
W(k)=\prod_{i=1}^k {1\over 1+b/i} ={k!\over (1+b)_k}={k!\,\Gamma (1+b)\over\Gamma (1+b+k)} ,
\eea
where $(a)_k =\prod_{i=0}^{k-1} (a+i)$ denotes the Pochhammer symbol, $a\in\rel$, $k\in\nat$. The grand canonical partition function of (\ref{grandcanz}) is
\bea\label{z}
Z(\phi )=\ _2 F_1 (1,1;1+b;\phi ):=\sum_{k=0}^\infty {(1)_k (1)_k \over (1+b)_k}{\phi^k \over k!}\ .
\eea
Its radius of convergence is $\phi_c =1$, and $_2 F_1$ denotes the hypergeometric function \cite{abramowitz}, which has the expansion
\bea
\lefteqn{_2 F_1 (k,k;k+b;\phi )={\Gamma (k+b)\Gamma (k-b)\over\Gamma (k)^2}\, (1-\phi )^{b-k} \Big[ 1+\ord (1-\phi )\Big] +}\nonumber\\
& &\qquad {\Gamma (k+b)\Gamma (b-k)\over\Gamma (b)^2} \left[ 1+{k^2\over 1+k-b} (1-\phi )+\ord (1-\phi )^2 \right]\ .
\eea
The particle density (\ref{rhodef}) is given by
\bea\label{rho}
\rho (\phi )={\phi\ _2 F_1 (2,2;2+b;\phi )\over (1+b)\, _2 F_1 (1,1;1+b;\phi )}\quad\mbox{for }\phi <1.
\eea

In the following we analyze the grand-canonical single site measure $\nu_\phi$ of (\ref{grandcanens}) in the limit $\phi\nearrow 1$, i.e. near the critical density $\rho_c$. For $\rho_c <\infty$ the limit $\nu_1$ is well defined, as discussed in Section 2, and according to Theorem 1 it is the distribution of the non-condensed phase for supercritical systems with $N/L=\rho >\rho_c$. As long as $\phi <1$ the distribution $\nu_\phi$ has exponential moments. For $\phi =1$ the exponential tail of $\nu_\phi$ disappears and the tail becomes proportional to the weight $W(k)$ as defined in (\ref{weights}). Using Stirling's formula, the behavior of the weight for large $k$ is given by the power-law $W(k)\simeq\Gamma (1+b)\,k^{-b}$. These distributions have moments up to order $b-1$. Thus different scenarios are encountered as $b$ is varied.\\
\\
\textbf{The case $0<b\leq 1$:}\\
For $b<1$ the leading order in the asymptotic expansion for $Z$ and $\rho$ is given by
\bea
Z(\phi )&\simeq &\Gamma (1+b)\Gamma (1-b)\, (1-\phi )^{b-1}\ \rightarrow\ \infty\ ,\nonumber\\
\rho (\phi )&\simeq &{\phi\over (1+b)^2 (1-b)}\, (1-\phi )^{-1}\ \rightarrow\ \rho_c =\infty\ ,
\eea
as $\phi\nearrow 1$. For every density, the stationary distribution in the limit $L\to\infty$ is given by the grand-canonical measure $\nu_\phi$ (see Figure \ref{phi}(a)). The probability to have a fixed number of particles on a given site vanishes with increasing density, as is shown for the example of an empty site in Figure \ref{phi}(b). Thus in the limit there is an infinite number of particles on every site with probability one, as it should be for homogeneous systems with $\rho\to\infty$.

For $b=1$ this picture does not change qualitatively, except for the logarithmic corrections
\bea
Z(\phi )&=&-{\log (1-\phi )\over\phi }\ \rightarrow\ \infty\ ,\nonumber\\
\rho (\phi )&=&{\phi\over (\phi -1)\log (1-\phi )}-1\ \rightarrow\ \rho_c =\infty\ ,
\eea
as $\phi\nearrow 1$.\\
\\
\bef
\begin{center}
\includegraphics[width=\textwidth]{phinu.epsi}
\end{center}
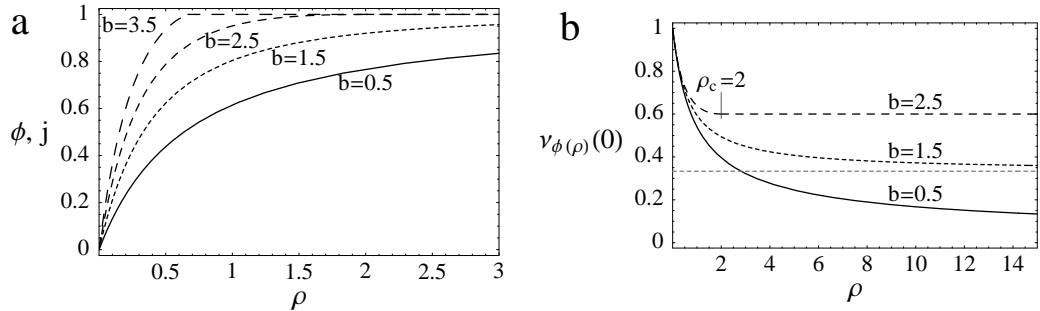
\caption{\small (a) Fugacity $\phi$ as a function of the particle density for several values of $b$. Equivalently, current $j$ as a function of $\rho$ up to the scale factor $m(p)$ (see (\ref{current})). (b) Probability of an empty site, $\nu_{\phi (\rho )}(0)$, as a function of the density $\rho$ for several values of $b$.}
\label{phi}
\enf
\\
\textbf{The case $1<b\leq 2$:}\\
For $1<b<2$ the leading order terms change, and
\bea
Z(\phi )&\simeq &{b\over b-1}+\Gamma (1+b)\Gamma (1-b)\, (1-\phi )^{b-1}\ \rightarrow\ Z(1)={b\over b-1}\ ,\nonumber\\
\rho (\phi )&\simeq &\phi (b-1)\Gamma (b)\Gamma (2-b)\, (1-\phi )^{b-2}\ \rightarrow\ \rho_c =\infty\ ,
\eea
as $\phi\nearrow 1$. As before, for $b=2$ the first order terms have logarithmic corrections but the qualitative behavior does not change. In particular, $\rho_c =\infty$ and the stationary distribution is described by the grand-canonical ensemble for every density $\rho$ (see Figure \ref{phi}(a)).

However, somewhat surprisingly, the character of this distribution for large $\rho$ differs from the case $b\leq 1$. Since $Z(1)<\infty$, $\nu_1$ is well defined and there is a non-zero probability to have a fixed number of particles at a given site,
\bea\label{finite}
\nu_1 (0)&=&{1\over Z(1)}={b-1\over b}\ ,\nonumber\\
\nu_1 (k)&=&{W(k)\over Z(1)}\approx\Gamma (b)\, (b-1)\, k^{-b}\quad\mbox{for large }k .
\eea
For example the probability of an empty site, given by $\nu_{\phi (\rho )} (0)=1/Z(\phi (\rho ))$, decreases monotonically with $\phi$, i.e.\ with increasing density $\rho$. In contrast to the case $b\leq 1$, it does not vanish in the limit $\rho\to\infty$, however, it reaches the non-zero value $\nu_1 (0)=(b-1)/b$ (see Figure \ref{phi}(b)). So no matter how large the density, the fraction of empty sites in a typical configuration is always greater than $(b-1)/b$.

Distributions with power-law tails are well studied (see e.g.\ \cite{bardouetal02} and references therein). A typical configuration for this stationary distribution, i.e.\ a set of $L$ i.i.d.\ random variables $\eta_x$ drawn from $\nu_1$, is known to have a hierarchical structure. The $n$-th largest value of the set $\{\eta_1 ,\ldots ,\eta_L\}$ scales as $(\Gamma(b-1) L/n)^{1/(b-1)}$, which holds for every $b>1$. In our particular case $1<b\leq 2$, this means that the particle number $\Sigma_L$ also scales as $L^{1/(b-1)}$ and thus grows faster than the number of summands $L$. Therefore the particle density $\Sigma_L /L$ diverges as $L^{(2-b)/(b-1)}$ and the highest occupied site contains a nonzero fraction of the particles in the system. This hierarchical structure of typical configurations can be understood as a precursor for the condensation phenomenon to be discussed in the next part.\\
\\
\textbf{The case $b>2$:}\\
In this case $b$ is large enough so that besides the normalization also the first moment of the grand canonical distribution converges in the limit $\phi\nearrow 1$:
\bea\label{cg2skal}
Z(\phi )&\simeq &{b\over b-1}-{b\over (b-1)(b-2)}\, (1-\phi )\ \rightarrow \ Z(1)= {b\over b-1}\ ,\nonumber\\
\rho (\phi )&\simeq &{1\over b-2}\! +\!\phi (b-1)\Gamma (b)\Gamma (2-b)(1-\phi )^{b-2}\ \rightarrow\ \rho_c \! =\!{1\over b-2}\ ,
\eea
as $\phi\nearrow 1$. We note that for $b>3$ also the second moment $\sigma^2$ of the distribution $\nu_1$ exists and the number of particles satisfies the usual central limit theorem
\bea\label{clt}
\lim_{L\to\infty }\nu_1^L \left(\xi_1 \leq {\Sigma_L -\rho_c L\over\sigma\sqrt{L} }\leq\xi_2 \right) =\int_{\xi_1}^{\xi_2} G(\xi )\, d\xi ,
\eea
where $G$ denotes the Gaussian probability density with zero mean and unit variance. The density for $b>3$ is of order $\rho (\phi )=1/(b-2) +\ord (1-\phi)$ and its first derivative is finite at $\phi =1$ and given by $\rho '(1)=(b-1)^2 /((b-3)^2 (b-2)^2 )$ (see Figure \ref{phi}(a)).

As explained already in Section 3, the most occupied site contains of order $L^{1/(b-1)}$ particles, and for $b<3$ this fluctuation is larger than $\sqrt{L}$. Therefore the scaling limit leads to a self-similar distribution, which is given by the completely asymmetric L\'evy distribution $L_{(b-1)}$ (for details see \cite{bardouetal02} or \cite{gnedenko}),
\bea\label{cltlevy}
\lim_{L\to\infty }\!\nu_1^L \!\!\left(\xi_1 \leq {\Sigma_L -\rho_c L\over \Big( b\,\Gamma (b-1)\, L\Big)^{1/(b-1)} }\leq\xi_2 \right)\! =\!\int_{\xi_1}^{\xi_2}\!\! L_{(b-1)} (\xi )\, d\xi .
\eea
With (\ref{cg2skal}) we have $\rho '(1)=\infty$ for $b<3$, leading to a differentiable function $\phi (\rho )$, as shown in Figure \ref{phi}(a).

\section{Dynamics of the condensation}
The stationary distribution investigated so far carries no information on the kinetics of the condensation. A natural set-up is to start with particles uniformly distributed at the supercritical density $\rho >\rho_c$. In the beginning the excess particles condense at a few random sites. Such a site containing many excess particles is called a cluster site. Thus there are several clusters which are essentially immobile as will be discussed below. On the remaining sites, called bulk sites, the distribution relaxes to $\nu_1$. With increasing time the larger clusters will gain particles at the expense of the smaller ones, causing some of the clusters to disappear. Eventually only a single cluster containing all excess particles survives, which is typical for the stationary distribution, as has been discussed already in Section 3.

In the following we will study the kinetics of condensation and its dependence on system parameters in detail. Compared to Section 4 the jump probabilities play an important role for the dynamical properties of the system. We will focus on nearest-neighbor jumps which are either totally asymmetric, i.e. particles only jump to the right with $p_a (y)=\delta (1,y)$, or symmetric, i.e. $p_s (y)=\big( \delta (-1,y) +\delta (1,y)\big) /2$. In contrast to previous sections we are not able to rigorously prove our statements, but use heuristic considerations which are corroborated through comparison with simulation data.

  \subsection{Cluster formation}
To distinguish cluster and bulk sites, we define a site $x\in\{ 1,\ldots ,L\}$ to be a cluster if it contains a macroscopic fraction of the excess particles $\eta_x >\alpha (\rho -\rho_c )L$. The prefactor $\alpha\in (0,1]$ is rather arbitrary but for simulations it is important that the clusters are well separated from the bulk fluctuations, which are of order $(\Gamma (b-1)L)^{1/(b-1)}$. Since fluctuations grow only sublinearly with $L$ this separation is clearly guaranteed in the limit $L\to\infty$, and for finite systems it holds for sufficiently large $L$ depending on $\alpha$ and $\rho$. In our simulations we choose $\alpha =1/40$, requiring system sizes of about 200 sites minimum for the values of $\rho$ and $b$ considered. Let $n(t)$ be the number of cluster sites at time $t$ and $m_i (t)$, $i=1,\ldots ,n(t)$ be the size of the $i$-th largest cluster, i.e. $m_1 (t)\geq \ldots\geq m_{n(t)} (t)$. These quantities depend also on the system parameters $b$, $L$, and $\rho$.

By definition a typical cluster has a size of order $\alpha (\rho -\rho_c )L$ and so there are of the order of $1/\alpha$ cluster sites. The time scale for the formation of such clusters is very roughly estimated as follows: $\ord (L)$ particles have to move a distance of order $L$ to form the cluster. So in the asymmetric case the time scale for cluster formation is $\ord (L^2 )$. The dependence on $\rho$ and $b$ is not so obvious, since the bulk has not yet relaxed to $\nu_1$ and the speed of particles still changes. In the symmetric case the time scale is $\ord (L^3 )$, since the particles diffuse without a drift.

In Figure \ref{formation}(a), in the totally asymmetric case, we plot the average number of particles in the condensed phase, normalized by the number of excess particles,
\bea\label{fraction}
f(t)={\sum_{i=1}^{n(t)} m_i (t)\over (\rho -\rho_c )L}\ .
\eea
The time axis is scaled proportional to $\tau_a =(\rho -\rho_c )^2 L^2 /b$ (cf.\ Section 5.2), since this choice gives the best data collapse when $\rho$ and $b$ are varied. On this time scale most excess particles become trapped in a cluster and the bulk relaxes to $\nu_1$. In Figure \ref{formation}(b) the average number of clusters is plotted as a function of time. The number of clusters grows for a short time and then starts decreasing again.
\bef
\begin{center}
\includegraphics[width=\textwidth]{formation.epsi}
\end{center}
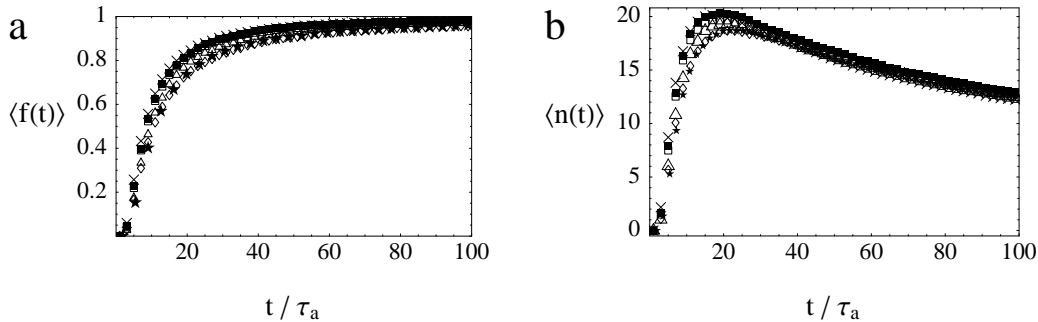
\caption{\small (a) Average fraction $\langle f(t)\rangle$ of particles in the condensed phase, (b) average number of cluster sites $\langle n(t)\rangle$. Both are plotted as a function of time in units $\tau_a =(\rho -\rho_c )^2 L^2 /b$ (cf.\ (\ref{timescales})) for different values of $\rho$, $b$ and $L$ in the totally asymmetric case. Symbols: for $b=4$, $\rho =5$, $L=320 (\Diamond )$, $640 (\triangle )$, $1280 (\Box )$, $2560 (\times )$, for $b=4$, $\rho =3$, $L=1280$ $(\bigstar )$, for $b=5$, $\rho =5$, $L=2560$ $(\blacksquare )$.}
\label{formation}
\enf

  \subsection{Coarsening}
Once the bulk has relaxed to $\nu_1$, each bulk site loses particles at the average rate $\langle g_b \rangle_{\nu_1} =1$. In the asymmetric case this results in a particle current $j=\phi =1$ (see (\ref{current}) and Figure \ref{phi}). On top of that, excess particles are exchanged between clusters. The bulk can be seen as a homogeneous medium where the excess particles move, and the cluster sites as boundaries where they enter and exit. A cluster of size $m>0$ loses excess particles with rate $g(m)-1=b/m$ and gains particles from neighboring clusters. Since this rate decreases with increasing cluster size, smaller clusters lose particles to the larger ones.

To quantitatively describe this coarsening process we study the normalized mean cluster size $\bar{m} (t)=f(t)/n(t)$ as a function of time for large system sizes $L$. The ensemble average of this quantity (denoted by $\langle\ldots\rangle$) is expected to grow according to a scaling law
\bea\label{scalinglaw}
\langle\bar{m} (t)\rangle\sim t^\beta\ ,
\eea
with a scaling exponent $\beta$ \cite{barmaetal02}. To estimate this exponent we notice that the time scale for the coarsening process is determined by two factors: Firstly, the rate at which excess particles leave a cluster of size $m$ and enter the bulk. Secondly, the typical time of such a particle to reach the neighboring cluster. The mobility of excess particles in the bulk is characterized by the average exit rate from an occupied site,
\bea\label{speed}
\langle g(k)|k>0\rangle_{\nu_1}\!\! =\!\sum_{k =1}^\infty g(k){\nu_1 (k)\over \nu_1 (k\neq 0)}\! =\!{\langle g(k)\rangle_{\nu_1} \over 1-\nu_1 (0)}\! =\!\Big(\! 1\! -\!{b\! -\! 1\over b}\Big)^{-1}\!\!\!\! =b\, ,
\eea
as follows from (\ref{current}) and (\ref{finite}). The excess particles perform a random walk in the bulk, which is either biased or unbiased, depending on the first moment of the jump probabilities. The time it takes to reach the next cluster at a distance of order $\ord (L/n)=\ord (m/(\rho -\rho_c ))$ is given by the mean first passage time for a random walk \cite{murthyetal89}. In the following we distinguish between asymmetric and symmetric jump probabilities, to estimate the time scale for a particle to leave a cluster and the one for reaching the neighboring cluster.

In the totally asymmetric case excess particles leave a cluster of size $m$ with rate $b/m$ and move towards the right neighboring cluster without returning, so the time to lose one particle scales like $\ord (m/b)$. Since the mean first passage time of a biased random walk is proportional to the distance, this particle spends a time $\ord (m/(b(\rho -\rho_c )))$ in the bulk, because $b$ is the speed of the particle. Thus the typical times for exiting a cluster and entering the next neighbor are of the same order in $L$ and there are $\ord (1/(\rho -\rho_c ))$ excess particles in the bulk. The coarsening time scale, determined by the typical time for a cluster to lose all $m$ particles, is thus proportional to
\bea\label{atime}
t_a (m):={m^2 \over b}\ ,\mbox{ i.e.}\quad\beta_a =1/2\ .
\eea
So for the asymmetric case the mean cluster size is predicted to grow like $\langle\bar{m} (t)\rangle\sim (bt)^{1/2}$.

In the case of symmetric jump probabilities $p_s$, excess particles perform an unbiased random walk in the bulk with diffusion constant $b$. Thus the mean first passage time to reach the next cluster is proportional to the square of the distance, i.e.\ $\ord (m^2 /(b\, (\rho -\rho_c )^2 ))$. In contrast to the asymmetric case, it is very likely that particles return to the cluster they left. The probability that they do not return but reach the neighboring cluster, which is the relevant event for coarsening, is inverse proportional to the diffusion distance, i.e.\ $\ord ((\rho -\rho_c )/m)$. So the typical time of a particle to leave a cluster is $\ord (m^2 /(b(\rho -\rho_c )))$ and as before there are $\ord (1/(\rho -\rho_c ))$ excess particles in the bulk. The coarsening time scale is therefore proportional to
\bea\label{stime}
t_s (m):= t_a (m)\, {m\over (\rho -\rho_c )}= {m^3 \over b(\rho -\rho_c )}\ ,\mbox{ i.e.}\quad\beta_s =1/3\ .
\eea
Thus the mean cluster size is predicted to grow like $\langle\bar{m} (t)\rangle\sim ((\rho -\rho_c )bt)^{1/3}$.

In general, the time scale for the coarsening regime is determined by the largest clusters with size $\ord ((\rho -\rho_c )L)$ and is thus of order 
\bea\label{timescales}
\tau_a &:=&t_a \big( (\rho -\rho_c )L\big) =(\rho -\rho_c )^2 L^2 /b\nonumber\\
\tau_s &:=&t_s \big( (\rho -\rho_c )L\big) =(\rho -\rho_c )^2 L^3 /b
\eea
for asymmetric resp.\ symmetric jump probabilities. The growth exponents for the two cases are confirmed by simulations and shown in Figure \ref{coars}(a) for the totally asymmetric and \ref{coars}(b) for the symmetric jump probabilities. Using the time scale $\tau_a$ resp.\ $\tau_s$ and normalizing $\langle\bar{m}\rangle$ by the number of excess particles $(\rho -\rho_c )L$ the data for different system sizes collapse. The measured growth exponents from these data are
\bea
\beta_a =0.514\pm 0.005\qquad\mbox{and}\qquad\beta_s =0.334\pm 0.004\ ,
\eea
which agree with the above predictions. Independently from us, these exponents have been obtained in \cite{godreche03} by numerical simulations.

Note that the clusters coarsen on the same time scale as they nucleate. However, looking at the number of time steps in Figures \ref{formation} and \ref{coars}, the time it takes to nucleate clusters is by a factor of 10 shorter than the coarsening regime.
\bef
\begin{center}
\includegraphics[width=\textwidth]{coars.epsi}
\hspace*{26mm} tot.\ asymmetric \hfill symmetric\hspace*{13mm}
\end{center}
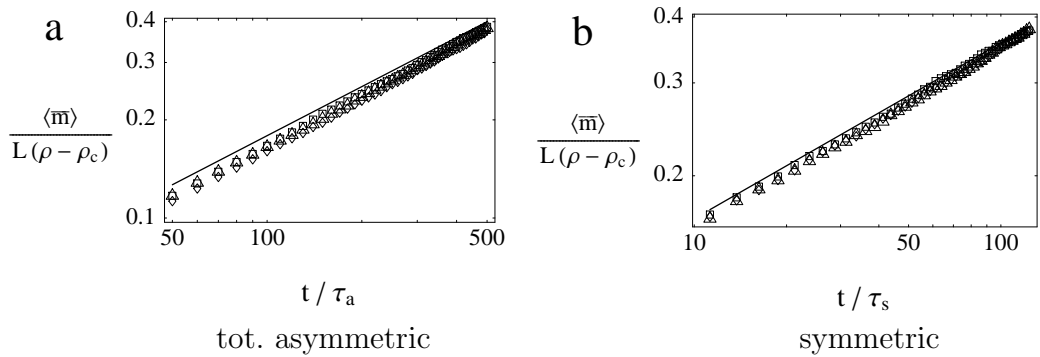
\caption{\small Double-logarithmic plot of mean cluster size as a function of time. Data collapse is achieved by using the appropriate time scales $\tau_a$, $\tau_s$ given in (\ref{timescales}), and dividing the cluster size by the number of excess particles $(\rho -\rho_c )L$. The straight line indicates the predicted slope, (a) $\beta =1/2$ in the asymmetric case, (b) $\beta =1/3$ in the symmetric case. Symbols: (a) $b=4$, $\rho =5$ and $L=320 (\Diamond )$, $640 (\triangle)$, $1280 (\Box )$; (b) $b=4$, $\rho =3$ and $L=160 (\Diamond )$, $320 (\triangle)$, $640 (\Box )$.}
\label{coars}
\enf

  \subsection{Saturation}
Eventually all clusters except for two will have disappeared and finite size effects become dominant. The scaling law (\ref{scalinglaw}) is no longer valid in this regime, since the mean cluster size $\bar{m}$ saturates towards its limiting value. The two clusters exchange particles until one of them vanishes and the system has reached its stationary state where all excess particles are concentrated at a single cluster site. In Figure \ref{saturate} we plot the average size of the three largest clusters $\langle m_i (t)\rangle$, $i=1,2,3$ normalized by $(\rho -\rho_c )L$. Note that the coarsening regime ends at latest when the third largest cluster has disappeared, and thus takes only about a tenth of the total equilibration time. In the following, we focus on the totally asymmetric jump probabilities, but the symmetric choice would lead to an effective evolution equation of the same form.
\bef
\begin{center}
\includegraphics[width=\textwidth]{saturate.epsi}
\hspace*{27mm} tot.\ asymmetric \hfill symmetric\hspace*{12mm}
\end{center}
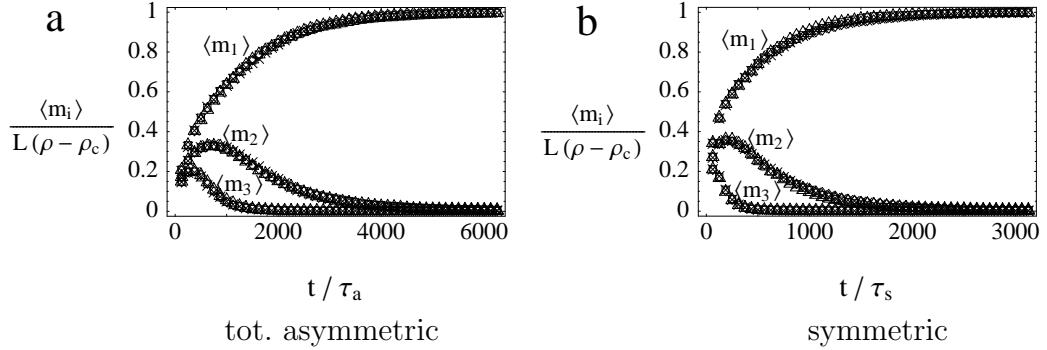
\caption{\small Average size of the three largest clusters $\langle m_i (t)\rangle$, $i=1,2,3$ as a function of time in the totally asymmetric case (a) and the symmetric case (b). Data collapse is achieved by using the appropriate time scales $\tau_a$, $\tau_s$ given in (\ref{timescales}), and dividing the cluster size by the number of excess particles $(\rho -\rho_c )L$. Symbols: (a) $b=4$, $\rho =20$ and $L=80 (\Diamond )$, $160 (\triangle)$, $\rho =40$ and $L=80 (\Box )$, $160 (\times )$. (b) $b=4$, $\rho =20$ and $L=40 (\Diamond )$, $80 (\triangle)$, $\rho =40$ and $L=40 (\Box )$, $80 (\times )$.}
\label{saturate}
\enf

Let $M=m_1 +m_2$ be the total number of particles at the two largest cluster sites. On the time scale $t^* =t\, /\, ((\rho -\rho_c )L/b)$ the two clusters exchange single particles with effective rates $(\rho -\rho_c )L/m_i$, $i=1,2$ (see discussion in Section 5.2). The fluctuations of $M$ on this time scale are only $\ord (1)$. Thus it is $M=(\rho -\rho_c )L+\ord (1)$, since the bulk is relaxed to $\nu_1$ and all other clusters have disappeared. Let $q(m,t^* )$ be the probability of having $m=0,\ldots ,M$ particles at one cluster site and $M-m$ at the other one. The dynamics is then governed by the effective master equation
\bea\label{master}
\lefteqn{{\partial\over\partial t^* }q(m,t^* )=-q(m,t^* )\left[{\theta (m)\over m/M}+{\theta (M-m)\over 1-m/M}\right]+\nonumber}\\
& &\qquad\qquad q(m\! -\! 1,t^* ){\theta (m)\over 1\! -\! (m\! -\! 1)/M}+q(m\! +\! 1,t^* ){\theta (M\! -\! m)\over (m+1)/M}\ .
\eea
The exchange rates on the right hand side only depend on the rescaled variable $m/M$ and not on the system parameters $\rho$, $L$ and $b$. For large $M$, $m/M$ varies on the time scale $t^* /\, M \simeq t\, /\, ((\rho -\rho_c )^2 L^2 /b)$, confirming that $\tau_a$ (\ref{timescales}) is the appropriate time scale for the two-cluster situation. Therefore the plots in Figure \ref{saturate} are independent of the system parameters. However, in the following discussion we stick to the time scale $t^*$ and the discrete variable $m$.

For any initial condition the solution of (\ref{master}) tends to the inverse binomial distribution $q^* (m)\sim 1\left/\left({M\atop m}\right)\right.$. It is symmetric around $m=M/2$ and for small $m$ we have $q^* (m)=\ord (M^{-m})$. Thus the two extreme occupation numbers $m=0$ resp.\ $m=M$ are the most probable ones and in the limit $L\to\infty$ both have probability $1/2$, consistent with the results of Section 3. For $m=\alpha M$, $\alpha\in (0,1)$ it is $q^* (m)=\ord \bigg(\sqrt{M} \Big(\alpha^\alpha (1-\alpha)^{1-\alpha}\Big)^M \bigg)$ using Stirling's formula. Thus, in the stationary state, the typical time for a macroscopic fluctuation of the cluster size diverges exponentially with the system size $L$.

To study the relaxation dynamics, we write (\ref{master}) in the canonical form, using the discrete derivative $\nabla_m \, f(m):=f(m+1)-f(m)$,
\bea\label{master2}
{\partial\over\partial t^* }q(m,t^* )&=&-\nabla_m \Big( a(m)\, q(m,t^* )\Big) +\nabla_m^2 \Big( d(m)\, q(m,t^* )\Big)\nonumber\\
a(m)&=&{1\over (1-m/M)}-{M\over m} ={2m/M-1\over m/M(1-m/M)}\nonumber\\
d(m)&=&{1\over 2}\left( {1\over(1-m/M)}+{M\over m}\right) ={1\over 2m/M(1-m/M)}\ .
\eea
For ease of notation, we take $0<m<M$ and ignore the boundary terms. Note that (\ref{master2}) is symmetric around $m=M/2$. It describes diffusive motion in a double well potential with drift $a(m)$ and diffusion coefficient $d(m)$, with the slightly unusual feature that the minima of the potential are located close to the boundaries at $m=1$ and $m=M-1$.

The master equation must be supplied with a suitable initial condition $q(m,0)$, which, since resulting from a complex coarsening process, is not readily available. A crude estimate can be found by noting that $q(m,0)$ is roughly proportional to the lifetime of the occupation number $m$. It is determined by the inverse exit rate for $m$ taken from Equation (\ref{master}), and thus we expect $q(m,0)\approx 6m/M\, (1-m/M)$. This is a symmetric single hump distribution with mean $M/2$ and standard deviation $M/(2\sqrt{5})\approx 0.22 M$. Comparing with the simulation data at the time when the third largest cluster has just disappeared, $m_3 (t)/M <0.01$, we indeed find a single hump distribution with mean $M/2$ and standard deviation $0.166 M$. When solving Equation (\ref{master2}) with this initial distribution, the expectation of the larger cluster size, given by $1/2+\langle |m/M-1/2|\rangle_{q(m,t)}$, is indistinguishable from $\langle m_1 (t)\rangle$ in Figure \ref{saturate}. Note that, except for the time scales, Figure \ref{saturate}(a) and \ref{saturate}(b) are almost identical, confirming that the effective master equation for the symmetric case is of the same form as (\ref{master}).

\section*{Acknowledgments}
S.G.\ acknowledges the support of the Graduiertenkolleg ``Mathematik im Bereich ihrer Wechselwirkung mit der Physik'' and the support of the\linebreak DAAD/CAPES program ``PROBRAL''. He is also grateful for a fruitful research visit at the ``Forschungszentrum J\"ulich'', where this work was initiated.


\end{document}